\documentclass[%
 prl,twocolumn,
 nofootinbib,
groupedaddress,
 amsmath,amssymb,
 aps,
]{revtex4-1}
\usepackage{amsmath}
\usepackage{graphicx,xcolor}
\usepackage{dcolumn}
\usepackage{bm}
\usepackage{hyperref}

\begin{document}


\title{Gravitational Wave Background Sky Maps from Advanced LIGO {\tt O1} Data}

\author{Arianna I. Renzini}
\email{arianna.renzini15@imperial.ac.uk}
\author{Carlo R. Contaldi}%
\affiliation{%
 Blackett Laboratory, Imperial College London, SW7 2AZ, UK}
 

\date{\today}

\begin{abstract}
We integrate the publicly available {\tt O1} LIGO time--domain data to obtain maximum--likelihood constraints on the Gravitational Wave Background (GWB) arising from stochastic, persistent signals. Our method produces sky--maps of the strain intensity $I$ as a function of direction on the sky at a reference frequency $f_0$. The data is integrated assuming a set of fixed power--law spectra for the signal. The maps provide upper limits on the amplitude of the GWB density $\Omega_{\rm GW}(f_0)$ and any anisotropy around the background. We find 95\% confidence upper limits of $\Omega_{\rm GW} < 4.8\times 10^{-7}$ at $f_0=50$ Hz with similar constraints on a dipole modulation for the inspiral--dominated stochastic background case. 
\end{abstract}

\pacs{Valid PACS appear here}
\maketitle

{\sl Introduction.--} 
The measurement of gravitational wave signals by the Laser Interferometer Gravitational wave Observatories (LIGO) and the Virgo Interferometer, emitted during the final phases of the merger of massive, compact objects \cite{Abbott2016a,Abbott2017,Abbott2017b,Abbott2017a,Abbott2017c} constitutes the ``tip of the iceberg'' in terms of the potential for astrophysical observations.  These detections are determined by a high signal-to-noise ratio and have yielded precise characterisation of the emitting merging systems. There will however be many more event signals in the detector time-streams that do not rise sufficiently above the noise to be detected, but which contribute to a stochastic background of gravitational waves \cite{Christensen2018}. Any detection of such a stochastic signal or indeed, any constraint on its amplitude, offers the possibility of discovery of new astrophysical or cosmological phenomena \cite{Regimbau2011,Caprini2018}. 

Upper bounds for the GWB have been obtained from the {\tt S4} and {\tt S5} LIGO data sets \cite{LIGOS4,LIGOS5,Abadie2011}, and more recently from Advanced LIGO's first observing run {\tt O1} \cite{LIGO2016,LIGO2016a}. These are the result of a variety of estimation methods, including estimation of directional limits and narrowband radiometer searches aimed at specific sky positions, such as the Galactic centre. These methods comprise both coherent and incoherent integration of the data. Different techniques are optimised for different background types, making them complementary.

\begin{figure*}[th]
\centering
\begin{minipage}{.32\textwidth}
  \centering
  \includegraphics[width=0.95\linewidth]{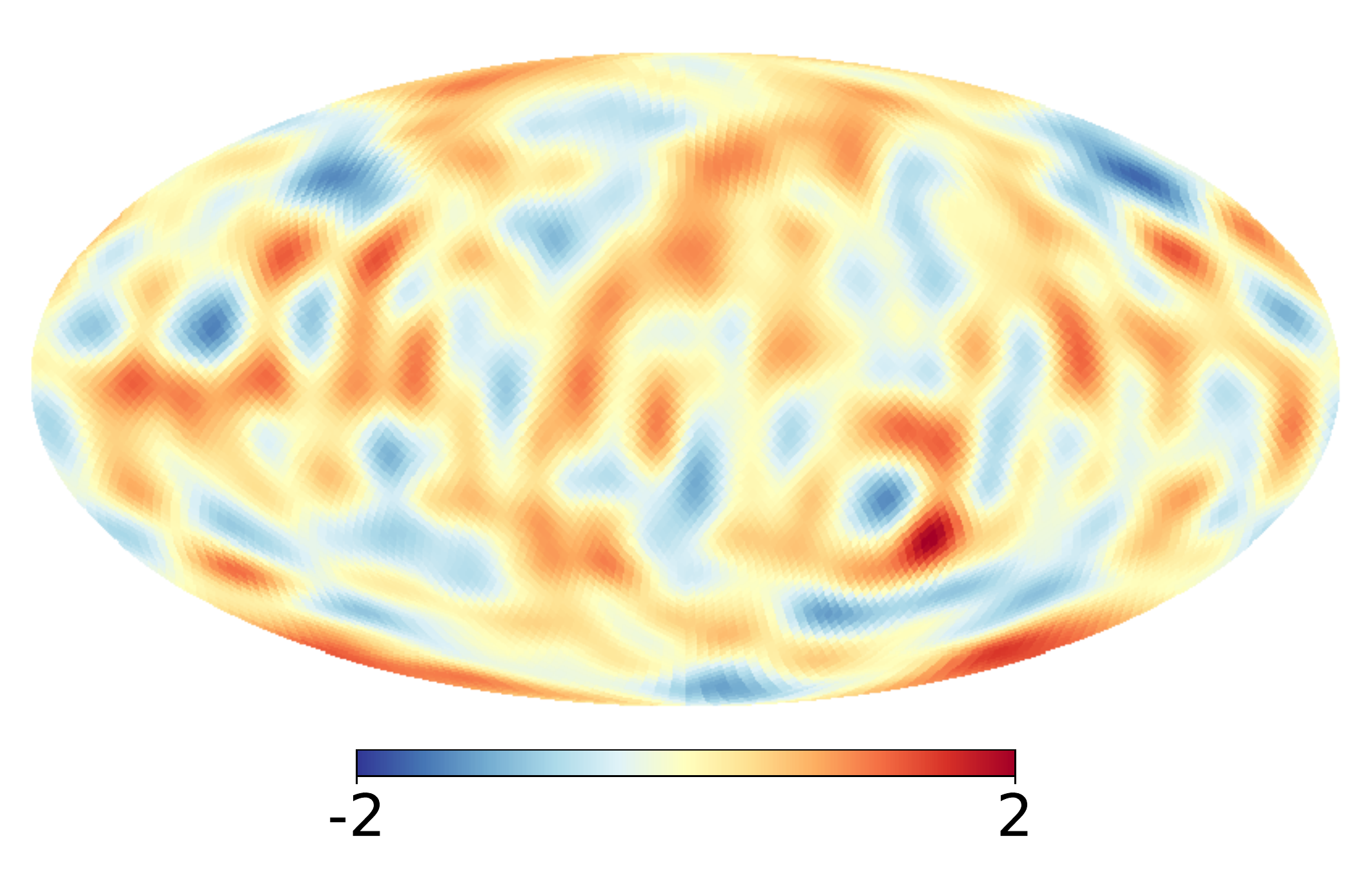}
\end{minipage}%
\begin{minipage}{.32\textwidth}
  \centering
  \includegraphics[width=0.95\linewidth]{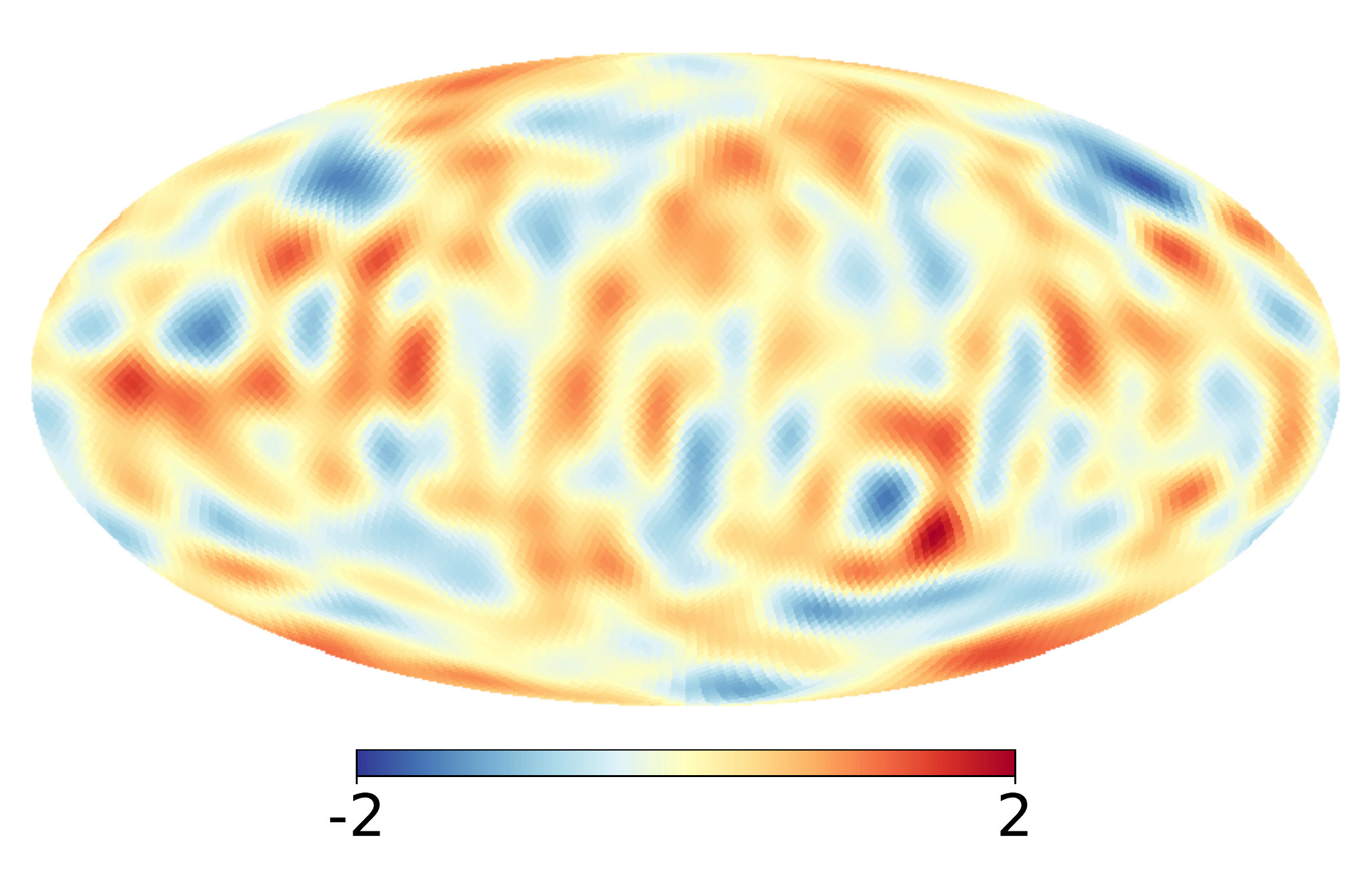}
\end{minipage}
\begin{minipage}{.32\textwidth}
  \centering
  \includegraphics[width=0.95\linewidth]{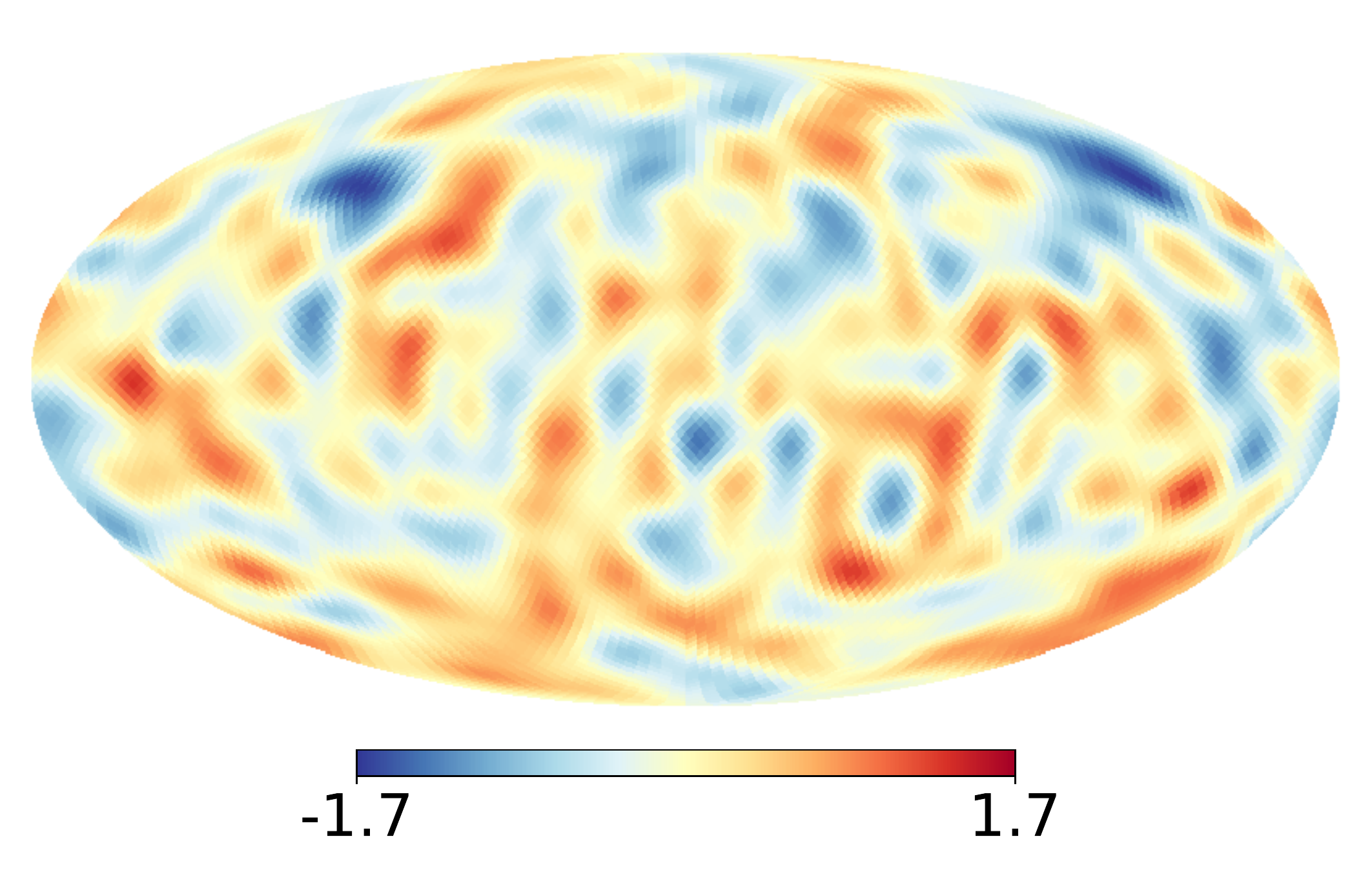}
\end{minipage}
\begin{minipage}{.32\textwidth}
  \centering
  \includegraphics[width=0.95\linewidth]{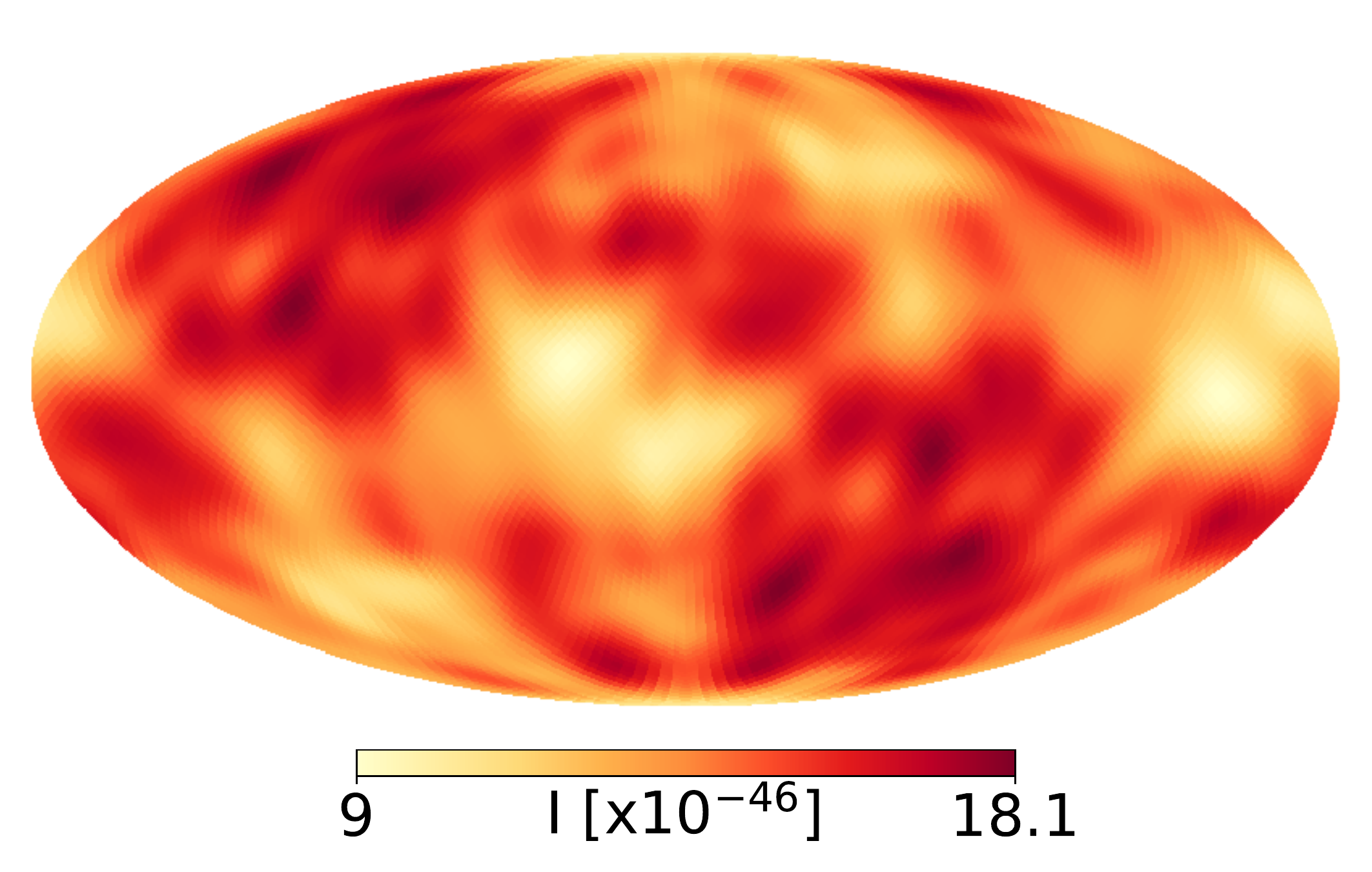}
\end{minipage}%
\begin{minipage}{.32\textwidth}
  \centering
  \includegraphics[width=0.95\linewidth]{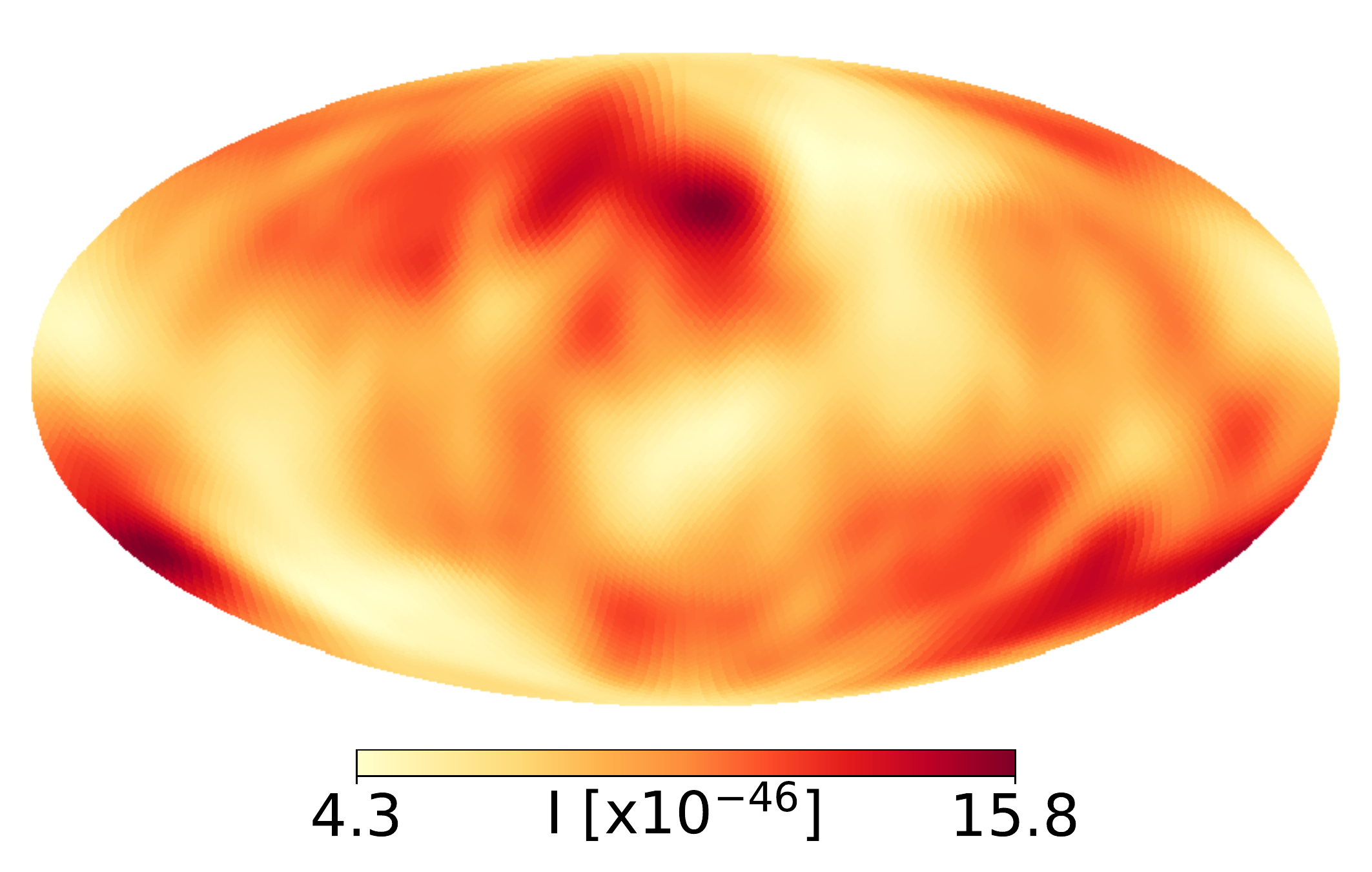}
\end{minipage}
\begin{minipage}{.32\textwidth}
  \centering
  \includegraphics[width=0.95\linewidth]{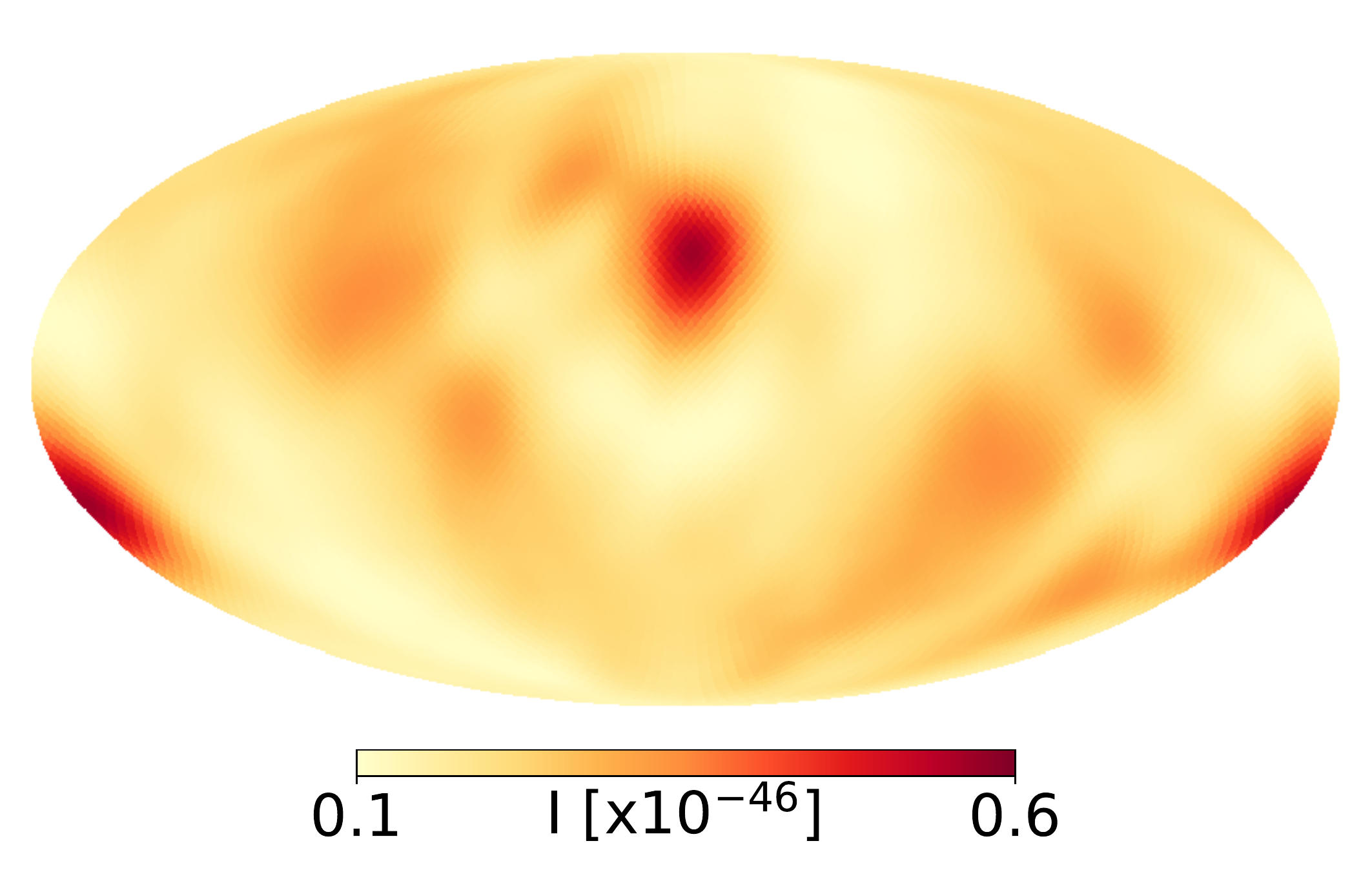}
\end{minipage}
\caption{SNR (\emph{top}) and noise maps (\emph{bottom}) of the for the intensity with, from left to right, $\alpha = 0,\,2/3,\,3,$ respectively. All maps have been produced at a {\tt HEALPix} resolution $N_{\rm side}=8$ corresponding to $n_{\rm pix}=768$. This corresponds to a pixelisation scale $\sim 7$ degrees or a Nyquist scale of $\ell\sim32$. For the purpose of visualisation we smooth the resulting maps with a 10 degree Gaussian beam and then over--resolve to $N_{\rm side}=32$. The $\alpha = 0,\,2/3,$ cases are at a reference frequency of $f_0 = 50$ Hz, whereas the $\alpha = 3$ case has $f_0 = 100$ Hz. The noise maps are in units of Hz$^{-1}$.
}
\label{fig:maps}
\end{figure*}

A method for obtaining maximum--likelihood sky--maps of strain intensity from an incoherent integration of general interferometric gravitational wave measurements was developed and tested in \cite{Renzini:2018vkx}. The method uses a generalised coordinate frame to obtain maps in galactic coordinates. Although the rotation to a generalised frame is an additional complication that is not strictly required when integrating the data from the single interferometric LIGO (Hanford--Livingston) baseline it will be an important component of any analysis using multiple baselines. The addition of coincident Virgo \cite{Abbott2017} detector data in the latest round of observations means we are already operating in the multiple baseline regime with more to come over the next decade. Space--based missions such as the Laser Interferometer Space Antenna (LISA) \cite{Amaro2017} will also see the generalised positioning of the detectors with respect to a fixed sky frame and this further justifies the development of true sky-mapping methods such as ours. 

In this {\sl letter} we report on the application of our map-making method to the LIGO {\tt O1} data set \footnote{https://www.gw-openscience.org/data} from the first Advanced LIGO observing run. The results presented here are the first which have been obtained independently of the LIGO and Virgo collaboration. These are also the first maps obtained by a direct inversion of the data onto the sky frame. This choice is different from any methods proposed beforehand to obtain maps with LIGO interferometers \cite{Romano2017}. In addition, we have built an independent data selection and processing pipeline which is distinct from any other used to analyse gravitational wave data.

The maps produced are of the {\sl total} strain intensity $I(\mathbf{\hat{n}})$ as a function of direction on the sky $\mathbf{\hat{n}}$ in the sense that they represent the intensity of the strain integrated over a range in frequency assuming a particular spectral distribution. At any point on the sky the quantity $I$ is related to the one-sided power spectral density as
\begin{equation}
I(\mathbf{\hat{n}})= \frac{1}{f_s}\int_{0}^{f_s} df\, I(f,\mathbf{\hat{n}}) \equiv \frac{1}{2f_s}\int_{0}^{f_s} df\, S_h(f)\,, 
\end{equation}
where $f$ is the frequency and $f_s$ is the maximum frequency to which the measurements are integrated. The quantity $I(f,\mathbf{\hat{n}})$ 
is a {\sl specific} intensity in units of Hz$^{-1}$. Note that we normalise the specific intensity to include a factor of $4\pi$ from an integration over the entire solid angle in order to obtain a quantity whose {\sl average} over the sky is related to the gravitational wave background energy density $\Omega_{\rm GW}$. This choice is dictated by the nature of the observations where the lack of a compact beam means that the measurement, at each discrete time interval, represents an integral of a response over the entire celestial sphere.

Throughout this work we assume a spectral distribution $E(f) = (f/f_0)^{\alpha-3}$ for the signal where $f_0$ is a reference frequency and the spectral index $\alpha$ takes on different values for various source mechanisms. Under this assumption we can regard the map $\hat I$ obtained from our maximum--likelihood estimator as being the total intensity at the reference frequency $f_0$.

The intensity can be related to the gravitational wave energy density in units of the critical density $\rho_c$ 
\begin{equation}
\Omega_{\rm GW}(f) = \frac{1}{\rho_c}\frac{d\rho_{\rm GW}}{d\ln f}\,,
\end{equation}
as a function of direction $\mathbf{\hat{n}}$ as \cite{Thrane2013}
\begin{equation}
\Omega_{\rm GW}(\mathbf{\hat{n}},f_0)= \frac{4\pi^2}{3H_0^2} f_0^{3}\,\hat I(\mathbf{\hat{n}},f_0)\,,
\label{omega_gw}
\end{equation}
where $H_0 = 70$ km s$^{-1}$ Mpc$^{-1}$ is the Hubble rate today. The result can be scaled to arbitrary frequencies using the assumed spectral dependence as $\Omega_{\rm GW}(f) = \Omega_{\rm GW}(f_0)(f/f_0)^\alpha$. 

Table~\ref{tab:spectra} summarises the spectral dependence of the intensity and background amplitude for the three common assumptions adopted for $E(f)$. These are a cosmological, scale invariant background of inflationary origin with $\alpha=0$ \cite{Caprini2018}, a background dominated by the confused signal from the inspiral of compact objects with $\alpha=2/3$ \cite{Sesana2008}, and a stochastic background with spectral index $\alpha=3$ \cite{Abadie2011}, which is the simplest phenomenological assumption.

\begin{table}[t]
\begin{center}
\caption{Spectral dependence for various source mechanisms assumed in this work.}\label{tab:spectra}
\begin{tabular}{ l | c |c | c |c }
\hline
 Source & $\alpha$ & $\Omega_{GW}$ & $I$ & $f_0$ [Hz]    \\
\hline
\hline
Cosmo & 0 & constant& $\sim f^{-3}$ & 50\\
Inspiral& 2/3 & $\sim f^{2/3}$ & $\sim f^{-7/3}$ &50 \\
Astro & 3 & $\sim f^{3}$ & constant & 100\\
\hline
\end{tabular}
\end{center}
\end{table}

The map--making algorithm used in this work is detailed in \cite{Renzini:2018vkx}. We only summarise here the methodology and  specific application  to the LIGO {\tt O1} data. The LIGO Livingston (LL) and LIGO Hanford (LH) Observatories record two separate time-streams. The time--domain data are released at a down-sampled frequency $f_s = 4096$ Hz in time stamped blocks of variable size along with quality flagging information \cite{Vallisneri2014}.  The {\tt O1} release covers the period of approximately  129 days from September 12, 2015 through to January 19, 2016. 

{\sl Methodology.--} Our pipeline identifies time-coincident data segments from both detectors and discards segments that don't pass a combination of quality flags. These include flagging due to intrinsic noise states of either detector and flagging of test inspiral or stochastic signal injection. This first set of data cuts reduces the total duration of the data to 49.3 days. The data is then segmented further into 60 second blocks and tapered using a narrow cosine window of width 3 seconds to reduce edge effects. Each 60 second time--stream segment is Fourier transformed to the frequency domain. All valid data segments are notch filtered in frequency to remove biases due to known harmonics \cite{Renzini:2018vkx} and are band passed in the frequency range $[30.0,500.0]$ Hz. 

A three parameter analytical model is fit to the power spectra of each segment in order to construct the optimal filter to weight the data segment-by-segment \cite{Renzini:2018vkx}. When the fitting indicates either of the spectra are not consistent with the model the segments are discarded. This generally indicates that the $1/f$ or $f$--tails in the  spectrum of the data deviate from a nominal form. This procedure cuts a further 3\% of the remaining data. 

Finally, each pair of coincident frequency domain segments $s_f^\tau$ (LL) and $r_f^\tau$ (LH) is cross-correlated to obtain a data vector $d_f^\tau = s_f^\tau r_f^{\tau\star}$, where $\tau$ labels the time segment.

The data is modelled in relation to a map of the signal on the sky $I_p$ as
\begin{equation}
d_f^\tau = \sum_p A^\tau_{fp} I_p + n^\tau_f\,,
\end{equation}
where $p$ is a pixel index for the map, and $A^\tau_{fp}$ is the observation operator that projects the sky signal into the baseline frequency domain for the pointing at time segment $\tau$. The term $n_f^\tau$ is a noise contribution assumed to be stationary over the time segment and characterised by the product of power spectra for each detector as $\langle n_f^\tau n^{\tau\star}_{f'}\rangle \equiv \delta(f-f') N_f = \delta(f-f') P_f^{\tau (s)}P_f^{\tau (r)}$. Given the expected signal--to--noise this is is treated as an unbiased estimate of the noise variance.

\begin{figure}[t]
\centering
\includegraphics[width=0.95\columnwidth]{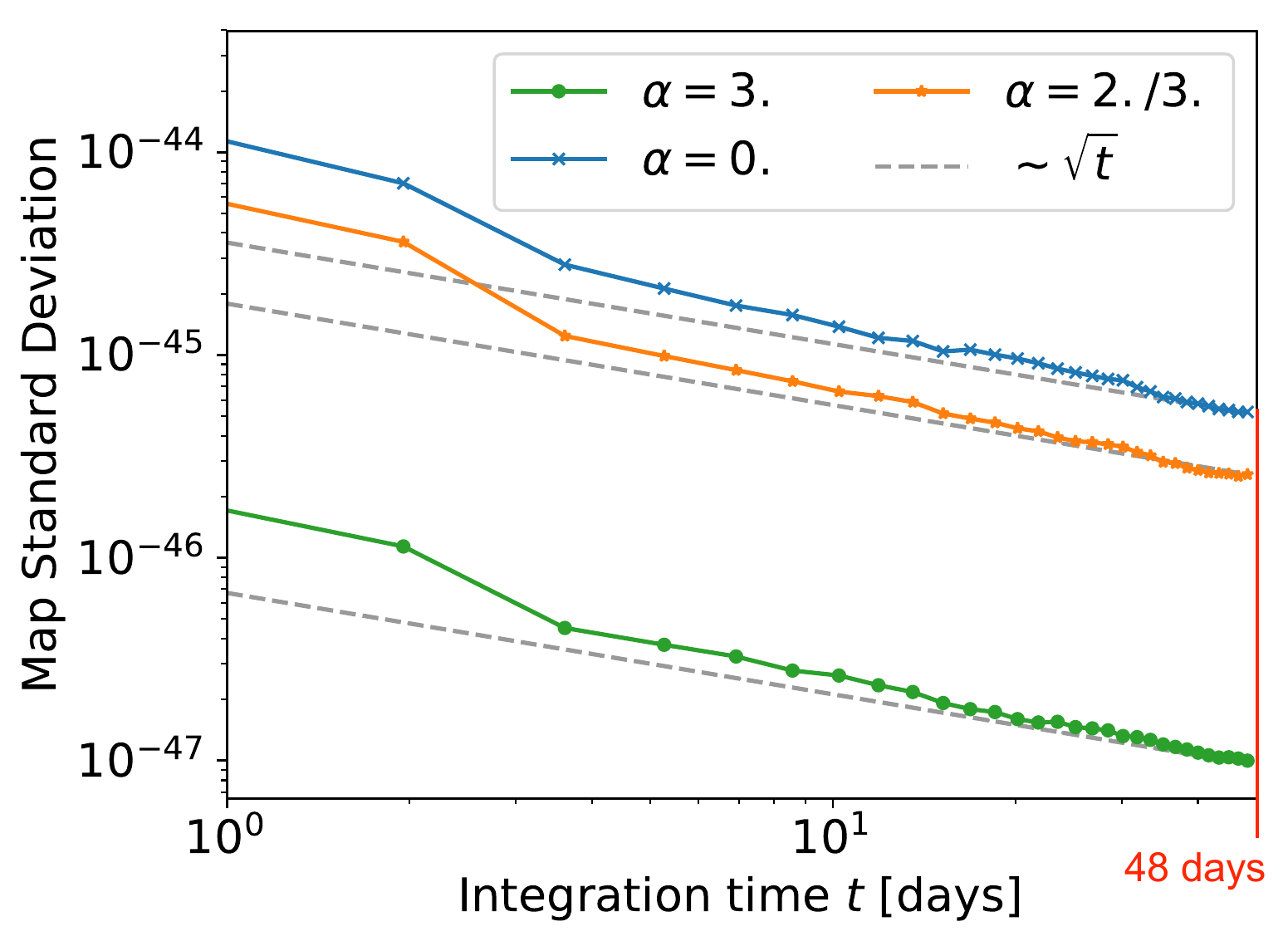}
\caption{Evolution of the standard deviation of the maximum--likelihood maps for the three spectral cases as a function of integration time. After a few days of integration the standard deviation enters a scaling regime proportional to the square root of the integration time consistent with the data being noise dominated and the solution being well conditioned.}
\label{fig:stats}
\end{figure}

The algorithm \cite{Renzini:2018vkx} accumulates the weighted map over all available segments $\tau$
\begin{equation}
z_p = \sum_{\tau, \,f\in \Delta f} A^\tau_{pf}\, N_f^{-1}\,d_f^\tau\,,
\end{equation}
and weight matrix
\begin{equation}
M_{pp'} = \sum_{\tau, \,f\in \Delta f} A^\tau_{pf}\, N_f^{-1} \, A^\tau_{fp'}\,,
\end{equation}
to obtain the maximum--likelihood estimate
\begin{equation}\label{eq:solution}
{\hat I}_p = \sum_{p'} M^{-1}_{pp'}\,z_{p'}\,.
\end{equation}
The projection operators $A^\tau_{fp}$ weight the frequencies by the assumed spectral dependence $E(f)$. As such the algorithm represents an optimal estimate for a signal with the given spectrum. For the present case, where the data is noise dominated, the variance in the estimated map will depend on the choice $E(f)$ or rather, the spectral index $\alpha$. Thus the maps will differ significantly for  different choices of $\alpha$ simply because the effective weights entering the integration of the data are significantly different as a function of frequency.

The inversion (\ref{eq:solution}) is more or less ill-conditioned for the single LIGO {\tt O1} baseline for a given choice of $\alpha$. To avoid numerical artefacts in the solution we carry out the inversion by using a pseudo--inverse method with conditioning threshold set to {\bf $10^{-5}$}. This nulls out any singular modes that represent modes on the sky that cannot be reconstructed using the current single baseline scan. We have found that the conditioning nulls out a small number of singular modes in the cosmological and inspiral--dominated spectrum cases.

The matrix $\mathbf{\cal N}\equiv\mathbf{M}^{-1}$ is the covariance of the maximum--likelihood map and can be used to assess the signal--to--noise (SNR) of the estimate. For the purpose of visualisation, since we do not expect a detection at the current sensitivity, we obtain SNR maps by calculating $\mathbf{s}=\mathbf{\cal N}^{-1/2} \, \mathbf{\hat I}$ where the matrix square root is obtained via diagonalisation. As the SNR maps are in units of the expected standard deviation they can be used to naively assess the significance of any feature.

\begin{figure}[t]
\centering
\includegraphics[width=0.95\columnwidth]{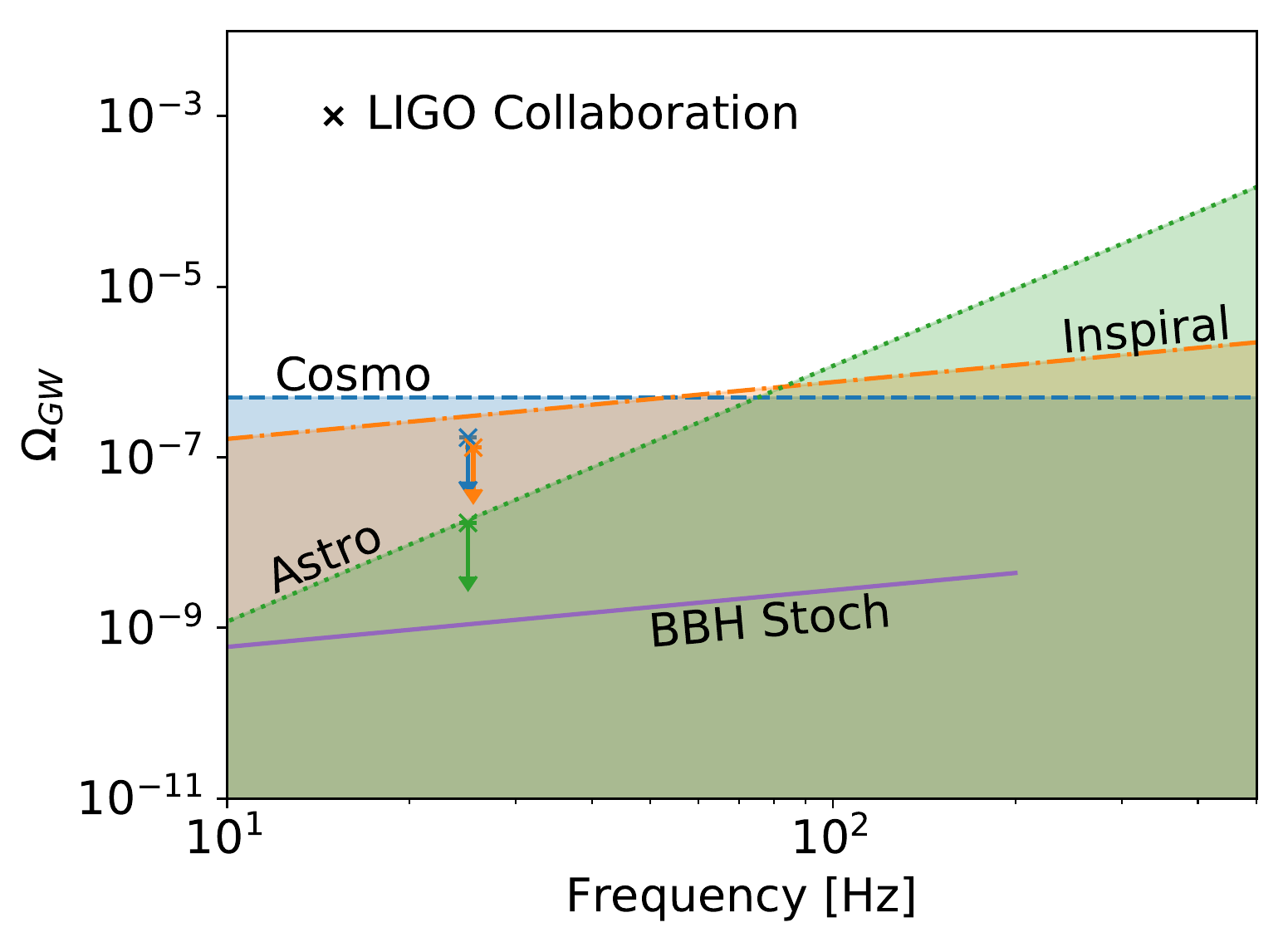}
\caption{95\% confidence, upper limits on $\Omega_{\rm GW}$ plotted as spectral functions for the three choices of $\alpha$. The arrows show corresponding upper limits reported in \cite{LIGO2016}, at $f_0=25$ Hz. The line labelled `BBH Stoch' corresponds to the approximate amplitude expected for the the stochastic inspiral signal calibrated off the detection rate of individual mergers observed so far \cite{LIGOimpli}.}
\label{fig:omegas}
\end{figure}

{\sl Results.--} We work at a {\tt HEALPix}\footnote{https://healpix.jpl.nasa.gov/}\cite{Gorski2004} resolution $N_{\rm side}=8$ when solving for the maps. In order to visualise the results we smooth the maps with a 10 degree Gaussian beam and re-sample to $N_{\rm side}=32$ to reduce noise at the pixelisation scale. Fig.~\ref{fig:maps} shows the SNR maps for all three cases $\alpha = 0$,  $2/3$, and $3$ considered in this work along with a map of the square root of the diagonal of the covariance $\mathbf{\cal N}$ for each case. The noise maps give an indication of the integration level across the sky given the scan of the single LIGO baseline. All maps are in galactic coordinates. The maps are consistent with noise in all three cases and scale with respect to integration time $t$ close to $t^{1/2}$, as seen in Fig.~\ref{fig:stats}. This is consistent with what we would expect in the noise dominated case if noise properties remain approximately constant throughout the run (after our quality cuts). The effective noise levels in the maps are different depending on the choice of spectral function $E(f)$. This is not surprising since the choice is based on an assumed {\sl signal} frequency dependence but the data is noise dominated. The most constraining case is $\alpha = 3$, as seen in the corresponding noise map of Fig.~\ref{fig:maps}. This is not surprising as this case results in the optimal weighting of the data with respect to the noise frequency dependence.

It is not straightforward to compare these results with the most recent directional limits from the LIGO and Virgo Collaboration presented in \cite{LIGO2016a} as the approaches are significantly different. Specifically, all maps presented here have been obtained at a fixed resolution determined by the working $N_{\text{side}}$, whereas the maps in \cite{LIGO2016a} are the result of a spherical harmonic decomposition with the effective resolution set by a cut-off in multipole at $\ell_{\text{max}}$ for each case. The significant correlations of the reconstructed modes in both methods mean that the visual appearance of the maps produced will be very different. However, the average value $\Omega_{\rm GW}$ should be consistent.

\begin{table}[t]
\begin{center}
\caption{Constraints on the isotropic background amplitude for different target spectral indices $\alpha$. The integration includes frequencies between 30 and 500 Hz.}

\label{tab:results}
\begin{tabular}{ l | c | c | c }
\hline
$\alpha$ & $f_0$  & $\Omega_{GW}$& 95\%  upper limit    \\
\hline
\hline
  0 & 50. &  $(2.4 \pm 2.5) \times 10^{-7}$ & $5.0\times 10^{-7}$ \\
2/3 & 50. & $(1.2 \pm 2.4)\times 10^{-7}$ & $4.8\times 10^{-7}$ \\
3 & 100. &$(1.1 \pm 5.9) \times 10^{-7}$ &  $1.2 \times 10^{-6}$ \\ 
\hline
\end{tabular}
\end{center}
\end{table}

We can extract upper limits, from our estimated maps, for the average value of $\Omega_{\rm GW}$ and any directional dependence on the sky from the maximum--likelihood maps $\mathbf{\hat{I}}$ using (\ref{omega_gw}). The limits for $\Omega_{\rm GW}$ are listed in table~\ref{tab:results} for the three distinct cases along with the choice of reference frequency at which the amplitude is evaluated. In addition we can obtain an estimate of any quantity $\mathbf{\hat{a}}$ obtained from the map by again using the maximum--likelihood solution
$\mathbf{\hat{a}} = (\mathbf{{ Y}}^\dagger \mathbf{{\cal N}}^{-1} \mathbf{{Y}})^{-1}\mathbf{{Y}}^\dagger \mathbf{{\cal N}}^{-1}\mathbf{\hat {I}}$ given a projection $  \mathbf{I}= \mathbf{{Y}}\,\mathbf{a}$. For example if $\mathbf{{ Y}}$ represents a spherical harmonic expansion and $\mathbf{a}$ are its coefficients we can obtain upper limits for the dipole components in $\mathbf{\hat {I}}$ and hence $\Omega_{\rm GW}$. These are of interest since our relative motion with respect to the cosmological rest frame guarantees the presence of a dipole of the order of 10$^{-3}\Omega_{\rm GW}$ in the presence of a uniform background. The limits for the three independent dipole components of $\Omega_{\rm GW}$ are shown in Table~\ref{tab:dipole}. Higher order multipoles may be even larger than this in the case of astrophysical sources \citep{Contaldi:2016koz,Cusin:2017fwz}

In Fig.~\ref{fig:omegas} we compare our upper limits for $\Omega_{\rm GW}$ with limits presented by the LIGO and Virgo Collaboration in \cite{LIGO2016}. We find they are in good agreement, especially in the case of $\alpha = 3$ which is also the optimal case for LIGO data. It would be clearly infeasible to integrate down to e.g. the expected stochastic signal shown in the figure. However, the constraint scales inversely with baseline noise and therefore planned upgrades to the LIGO detectors means levels of $\sim 10^{-9}$ should be within reach in the near future.

This research was supported by Science and Technology Facilities Council consolidated grants ST/L00044X/1 and ST/P000762/1. AIR acknowledges support of an Imperial College Schr\"{o}dinger Fellowship.

\begin{table}[t!]
\begin{center}
\caption{Constraints on the dipole components of the GWB for the three spectral indices. The corresponding reference frequencies are reported in Table~\ref{tab:results}.}
\label{tab:dipole}

\begin{tabular}{l |  c|  c| c}
\hline
 $\alpha$ &   $\hat{a}_{10}\, [\times 10^{7}]$& Re($\hat{a}_{11}$) $[\times 10^{7}]$ & Im($\hat{a}_{11}$) $[\times 10^{7}]$  \\
\hline
\hline
 $0$ & $-2.7 \pm 9.4$ &$2.1\pm 9.8$ &$-6.2 \pm 9.8$\\
$2/3$ & $-7.8 \pm 8.9$ &$1.3 \pm 9.3$ &$-3.7 \pm 9.3$\\
$3$  &$-41 \pm 20$ & $3 \pm 22$& $-6 \pm 22$ \\
\hline\hline
\end{tabular}
\end{center}
\end{table}

\bibliographystyle{apsrev}
\bibliography{refs.bib}
\end{document}